\documentclass{PoS}

\usepackage{xspace}

\title{Mixing and CP-violation studies in charm decays at LHCb}

\ShortTitle{Mixing and CP-violation studies in charm decays at LHCb}

\author{\speaker{Matthew CHARLES}\\
       On behalf of the LHCb collaboration\\
       The University of Oxford\\
       E-mail: \email{m.charles1@physics.ox.ac.uk}}

\abstract{Studies of charm physics with the 2010 LHCb data sample are presented.
  Time-integrated searches for CP violation in $D^+ \to K^- K^+ \pi^+$ and
  $D^0 \to K^- K^+, ~ \pi^- \pi^+$ are discussed.
}

\FullConference{The 2011 Europhysics Conference on High Energy Physics, EPS-HEP 2011,\\
		July 21-27, 2011\\
		Grenoble, Rh\^one-Alpes, France}

\def\CP {\ensuremath{C\!P}\xspace}
\def\AP {\ensuremath{A_{\mathrm{P}}}\xspace}

\def\AD {\ensuremath{A_{\mathrm{D}}}\xspace}
\def\pis {\ensuremath{\pi_{\mathrm{s}}}\xspace}
\def\aindCP {\ensuremath{a^{\mathrm{ind}}_{\CP}}\xspace}
\def\adirCP {\ensuremath{a^{\mathrm{dir}}_{\CP}}\xspace}
\def\ARAW {\ensuremath{A_{\mathrm{RAW}}}\xspace}
\def\invpb {\ensuremath{\mbox{\,pb}^{-1}}\xspace}

\def\Dbar {\kern 0.2em\overline{\kern -0.2em D}{}\xspace}

\def\Dz {\ensuremath{D^0}\xspace}
\def\Dzb {\ensuremath{\Dbar^0}\xspace}
\def\Dstarp  {\ensuremath{D^{*+}}\xspace}

\def\Bs      {\ensuremath{B^0_s}\xspace}
\def\Bbar    {\kern 0.18em\overline{\kern -0.18em B}{}\xspace}
\def\Bsb     {\ensuremath{\Bbar^0_s}\xspace}

\begin{document}

\section{Introduction}
\label{sec:intro}

The charm sector is a promising place to  probe for new physics
effects. Mixing
is now well-established~\cite{bib:hfag} at a level which is consistent
with but at the upper end of Standard Model (SM)
expectations~\cite{falk_grossman_ligeti_nir_petrov}. 
Three types of \CP violation (CPV) are possible:
  in the decay amplitudes,
  in the mixing between \Dz and \Dzb,
  and in the interference between mixing and decay.
The first is referred to as direct CPV, and the second and third as
indirect CPV. Only direct CPV is possible in $D^+$ decays, due to the absence of mixing.
In the SM indirect \CP violation is expected to be small
and direct \CP violation in singly-Cabibbo-suppressed modes such as
those discussed below is naively expected to be 
$\mathcal{O}(10^{-3})$ or less~\cite{bib:theory},
though larger values cannot be excluded from first principles~\cite{bib:kagan_dacp_2011}.
In the presence of new physics the rate of \CP violation
could plausibly be enhanced to $\mathcal{O}(10^{-2})$.
At the time of the conference no evidence for CPV
in charm had yet been found, though first indications have
since emerged in the 2011 LHCb data~\cite{bib:2011_dacp}.

\section{Search for CPV in $D^+ \to K^- K^+ \pi^+$}
\label{sec:KKP}

Direct \CP violation arises when two different amplitudes with
non-zero relative weak and strong phases contribute to decays to the
same final state. In two-body decays this must imply contributions from
different Feynmann diagrams, such as from tree and penguin processes.
In multi-body decays the same mechanism exists, but in addition a
rich variety of intermediate resonant states can contribute to the decay,
each naturally producing a different strong phase with well-defined
variation across the Dalitz plane. Thus, the interference between these
amplitudes can give rise to observable asymmetries which change across the 
Dalitz plane.

\begin{figure}
\begin{center}
  \includegraphics[width=.32\textwidth]{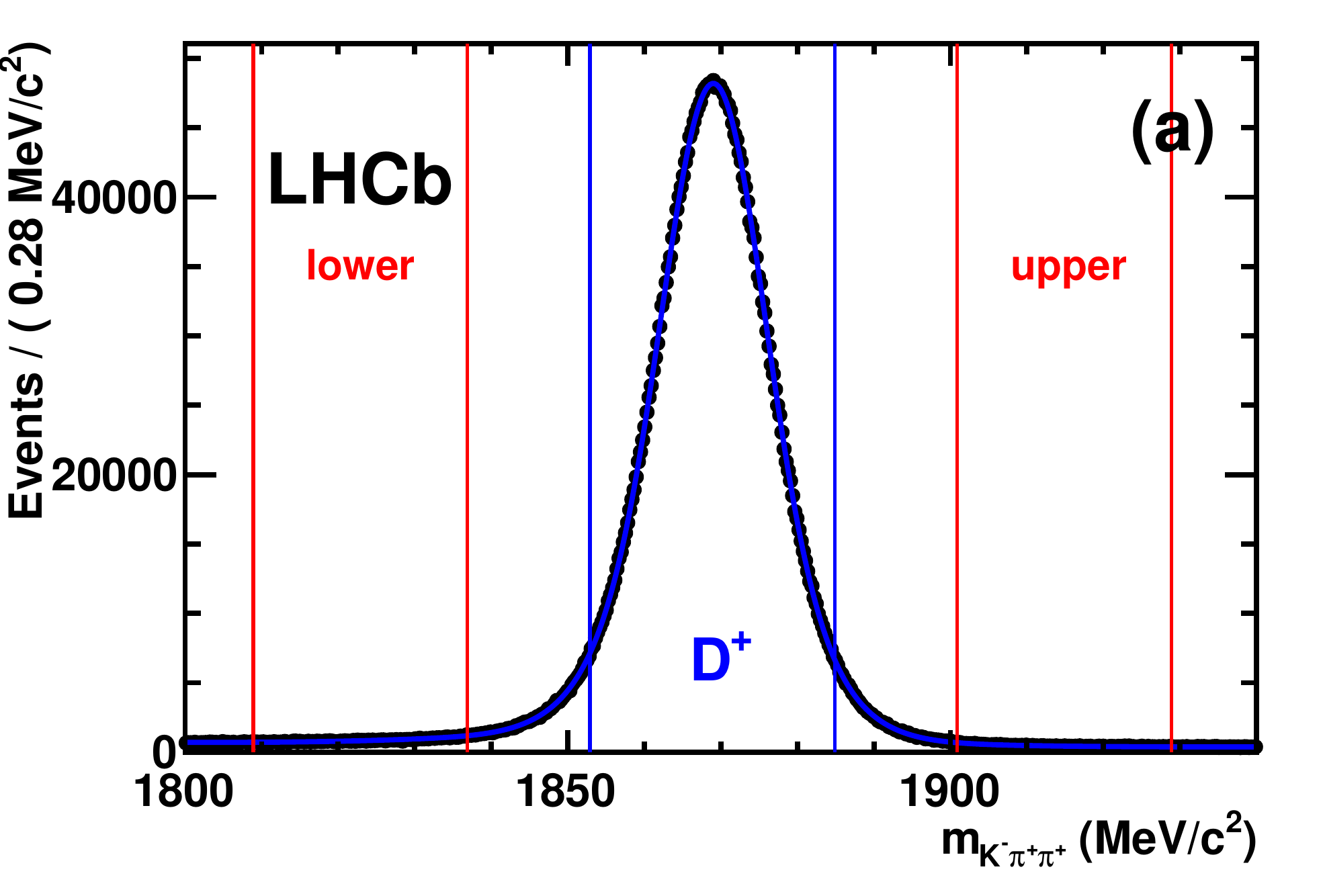}
  \includegraphics[width=.32\textwidth]{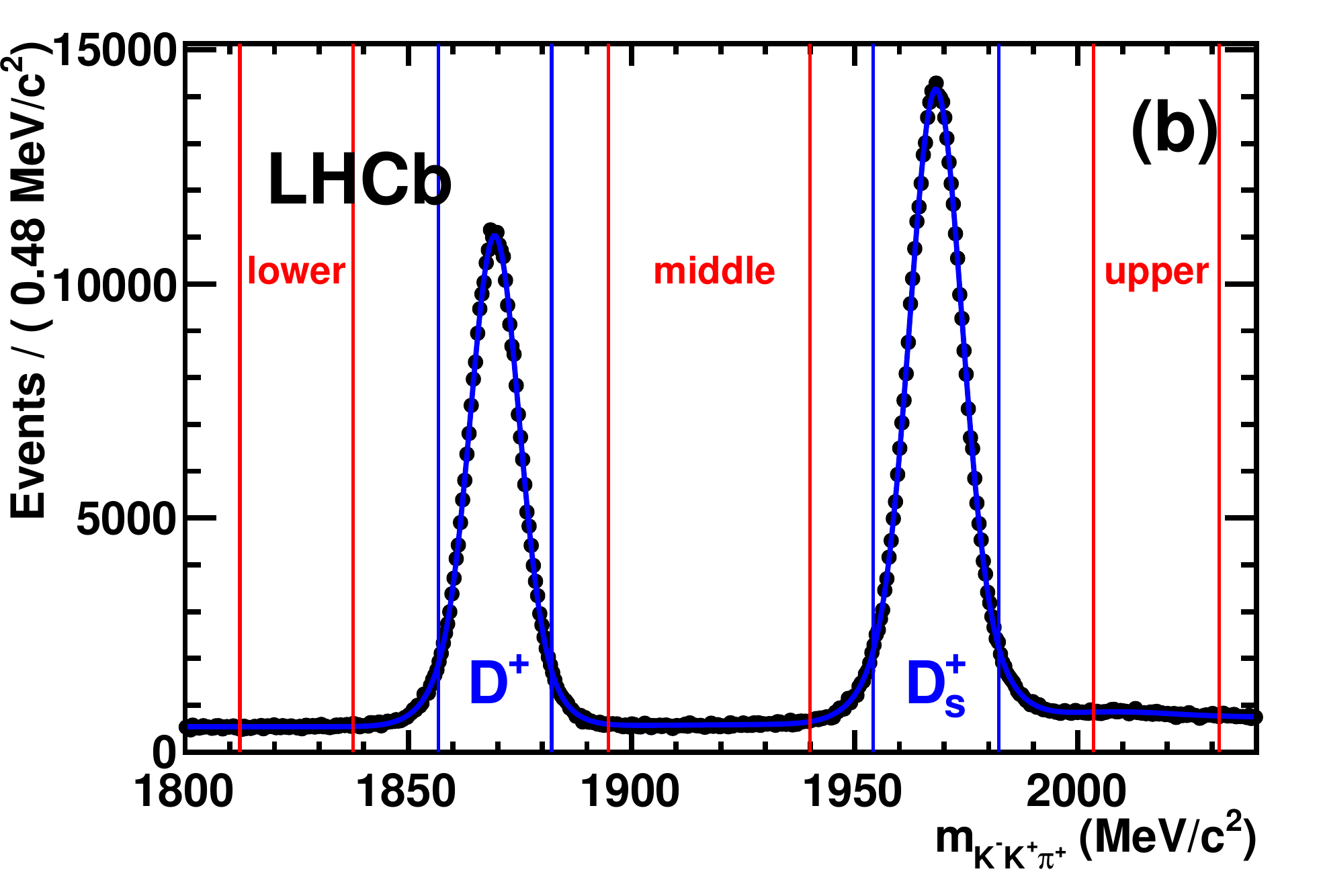}
  \includegraphics[width=.32\textwidth]{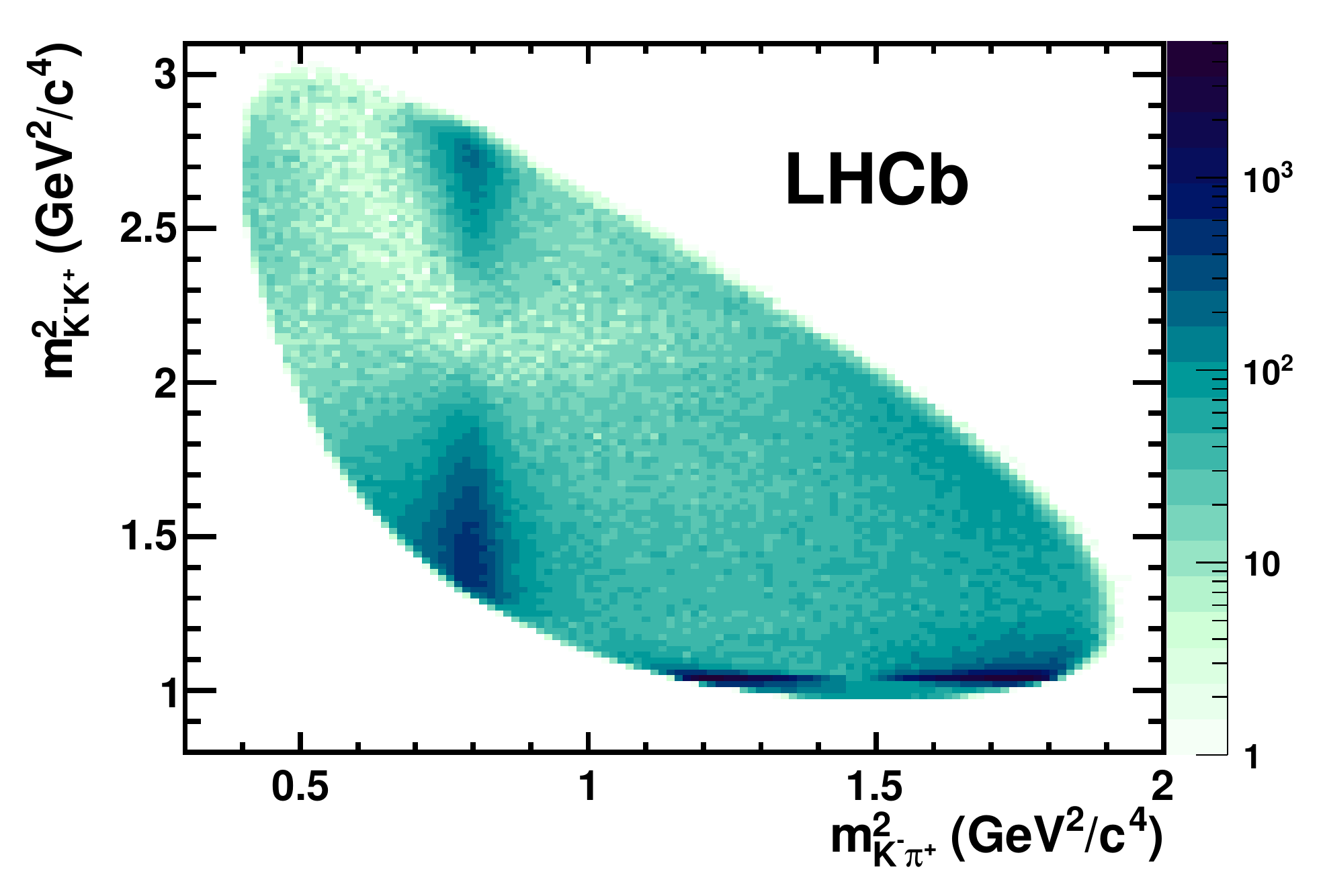}
\end{center}
\caption{
  Mass spectra and Dalitz plot.
  The mass spectra after selection for
  (a)~$K^- \pi^+ \pi^+$ and (b)~$K^- K^+ \pi^+$ are shown,
  with the signal and sideband mass windows indicated.
  For those candidates in the $D^+ \to K^- K^+ \pi^+$ signal
  window, the Dalitz plot is shown on the right.
}
\label{fig:KKP}
\end{figure}

We search for such asymmetries at LHCb~\cite{bib:LHCb} by comparing the Dalitz plot
distributions of $D^+ \to K^- K^+ \pi^+$ and its conjugate process
$D^- \to K^+ K^- \pi^-$ (Fig.~\ref{fig:KKP}),
applying a model-independent technique of
comparing the binned, normalized distributions. Normalizing the
two Dalitz plots to the same total number of events cancels
any production asymmetry and suppresses many systematic effects
that are mainly expressed as an overall efficiency asymmetry.
The statistical technique used to test for consistency between
the $D^+$ and $D^-$ Dalitz plots, and to localize the asymmetry
if one is found, is based on the Miranda approach
(see Ref.~\cite{bib:miranda} and also Ref.~\cite{bib:kalanand}).
A variety of different binnings are used
in order to test for different manifestations of $CP$ violation.

Control modes are analysed to validate the method. The main tool is the
Cabibbo-favoured $D_s^+ \to K^- K^+ \pi^+$ control mode,
which has the same final state as the signal as well as
similar kinematics and Dalitz plot structure. As expected,
no evidence of any asymmetry is found in this mode
(e.g. p-value of 34\% for 25-bin adaptive binning),
nor in the sidebands around the $D^+$ mass window.
In addition, the analysis is repeated for the Cabibbo-favoured
$D^+ \to K^- \pi^+ \pi^+$ mode. This is more sensitive to
systematic effects, since
  (a) the yield is ten times larger than that of the signal mode, and
  (b) the kaon imbalance can induce momentum-dependent detector
      efficiency asymmetries which would not be present in the signal mode.
Nonetheless, only weak indications of asymmetries are seen
(e.g. p-value of 12\% for 25-bin adaptive binning). Thus,
systematic effects in the more robust $D^+ \to K^- K^+ \pi^+$ signal
mode are negligible. The final, unblinded results are shown in
Table~\ref{tab:KKP}: no evidence of $CP$ violation is found in
the 2010 data. For further details, 
see Ref.~\cite{bib:paperKKP,bib:2011_dacp}.

\begin{table}
\begin{center}
\begin{tabular}{cccc}
\hline\hline
  Binning    & Bins & $\chi^2/{\rm ndf}$  &   $p$-value (\%)  \\ \hline
 Adaptive I  &   25 &        32.0/24      & 12.7 \\
 Adaptive II &  106 &       123.4/105     & 10.6    \\ 
 Uniform I   &  199 &       191.3/198     & 82.1   \\   
 Uniform II  &  530 &       519.5/529     & 60.5  \\     
\hline\hline
\end{tabular}
\end{center}
\caption{$\chi^2/{\rm ndf}$ and $p$-values
  for consistency with no CPV for the $D^+ \to K^- K^+ \pi^+$ decay
  mode with four different binnings.
}
\label{tab:KKP}
\end{table}

\section{Search for CPV in $D^0 \to K^- K^+, ~ \pi^- \pi^+$}
\label{sec:hh}

As discussed in Section~\ref{sec:intro}, both direct and indirect
CPV can contributed to the time-integrated $CP$ asymmetry in 
these singly Cabibbo suppressed decays to $CP$-even final states.
The indirect $CP$ asymmetry is universal to a very good
approximation~\cite{bib:sokoloff}, although the measured value
is affected by the \Dz decay time acceptance of the
experiment~\cite{bib:cdf_dacp}.
However, the direct $CP$ asymmetry in general varies between final
states, and in the limit of U-spin symmetry is equal and opposite
between $K^-K^+$ and $\pi^-\pi^+$~\cite{bib:theory}.
Thus, the difference in time-integrated asymmetry between
the two final states, $\Delta A_{CP}$, is sensitive to direct CPV
but has limited sensitivity to indirect CPV:
\begin{displaymath}
  \Delta A_{CP} = \adirCP(K^- K^+)
  - \adirCP(\pi^- \pi^+)
  + \frac {\Delta \langle t \rangle}{\tau} \, \aindCP,
\end{displaymath}
where $\frac {\Delta \langle t \rangle}{\tau} = 0.10 \pm 0.01$
is the 
difference in normalized time acceptance for the two
final states at LHCb, $\adirCP(f)$ is the direct $CP$
asymmetry for final state $f$, and $\aindCP$ is the indirect $CP$ asymmetry.

The observable $\Delta A_{CP}$ also has the advantage of being highly
robust against systematic effects. The measured (raw) asymmetry between
$\Dz \to f$ and $\Dzb \to \bar{f}$, where the initial flavour of the
$D$ is established with a $\Dstarp \to \Dz \pis$ tag, can be written at
first order as:
\begin{equation}
\ARAW(f) \approx A_{\CP}(f) \, + \, \AD(f) \, + \, \AD(\pis) \, + \, \AP(D^{*+}),
\label{def:arawstarcomponents}
\end{equation}
where $A_{CP}$, $\AD$, and $\AP$ are the relevant
physics, detector efficiency, and production asymmetries, respectively.
Within a local kinematic region, $\AD(\pis)$ and $\AP(\Dstarp)$ are
independent of the \Dz decay mode and thus cancel in the difference
$\Delta A_{CP}$. Further, $\AD(K^-K^+)$ and $\AD(\pi^-\pi^+)$ are
zero by construction, since the final state is spinless and self-conjugate.
Thus, all detector and production effects cancel in $\Delta A_{CP}$
at first order. To ensure good behaviour at second order, the data are
divided into 12 disjoint kinematic bins, as well as being partitioned
according to trigger conditions and magnetic field polarity. Taking the
weighted average of the individual measurements, we obtain
$\Delta A_{CP} = (-0.28 \pm 0.70 \pm 0.25)\%$, where the first uncertainty
is statistical and the second is systematic (taking into account
  modeling of the lineshapes [0.06\%],
  the \Dz mass window [0.20\%],
  multiple candidates [0.13\%], and
  the kinematic binning [0.01\%]).
For further details, see Ref.~\cite{bib:confDACP}.

\section{Conclusions and prospects}

LHCb's charm physics programme is off to a strong start.
Several proof-of-concept measurements have been made on the
2010 data sample of 38~\invpb,
and the first results on the much larger 2011 and 2012 data sets
are now forthcoming.


\begin{thebibliography}{99}

\bibitem{bib:hfag}
  Heavy Flavor Averaging Group,
  D.~Asner {\it et al.},
  {\it Averages of $b$-hadron, $c$-hadron, and $\tau$-lepton Properties},
  arXiv:1010.1589.

\bibitem{falk_grossman_ligeti_nir_petrov}
  A.~F.~Falk, Y.~Grossman, Z.~Ligeti, Y.~Nir and A.~A.~Petrov,
  {\it The \Dz-\Dzb mass difference from a dispersion relation},
  Phys.\ Rev.\  D {\bf 69} (2004) 114021,
  [hep-ph/0402204].

\bibitem{bib:theory}
  See e.g.
  S.~Bianco, F.~L.~Fabbri, D.~Benson and I.~Bigi,
  {\it A Cicerone for the physics of charm},
  Riv.\ Nuovo Cim.\  {\bf 26N7} (2003) 1
  [hep-ex/0309021];
  M.~Bobrowski, A.~Lenz, J.~Riedl and J.~Rohrwild,
  {\it How large can the SM contribution to $CP$ violation in \Dz-\Dzb mixing be?},
  JHEP {\bf 1003} (2010) 009
  [arXiv:1002.4794];
  Y.~Grossman, A.~L.~Kagan and Y.~Nir,
  {\it New physics and $CP$ violation in singly Cabibbo suppressed $D$ decays},
  Phys.\ Rev.\  D {\bf 75} (2007) 036008
  [hep-ph/0609178].

\bibitem{bib:kagan_dacp_2011}
  J.~Brod, A.~L.~Kagan and J.~Zupan,
  {\it On the size of direct $CP$ violation in singly Cabibbo-suppressed $D$ decays},
  arXiv:1111.5000.

\bibitem{bib:2011_dacp}
  LHCb Collaboration,
  R.~Aaij {\it et al.},
  {\it Evidence for $CP$ violation in time-integrated $\Dz \to h^-h^+$ decay rates},
  arXiv:1112.0938 (submitted to Phys.\ Rev.\ Lett.).

\bibitem{bib:LHCb}
  LHCb Collaboration,
  A.~A.~Alves, Jr. {\it et al.},
  {\it The LHCb Detector at the LHC},
  JINST {\bf 3} (2008) S08005.

\bibitem{bib:miranda}
  I.~Bediaga, I.~I.~Bigi, A.~Gomes, G.~Guerrer, J.~Miranda and A.~C.~d.~Reis,
  {\it On a $CP$ anisotropy measurement in the Dalitz plot},
  Phys.\ Rev.\ D {\bf 80} (2009) 096006
  [arXiv:0905.4233].

\bibitem{bib:kalanand}
  BABAR Collaboration,
  B.~Aubert {\it et al.},
  {\it Search for $CP$ Violation in Neutral $D$ Meson Cabibbo-suppressed Three-body Decays},
  Phys.\ Rev.\ D {\bf 78} (2008) 051102
  [arXiv:0802.4035].

\bibitem{bib:paperKKP}
  LHCb Collaboration,
  R.~Aaij {\it et al.},
  {\it Search for CP violation in $D^{+} \to K^{-}K^{+}\pi^{+}$ decays},
  arXiv:1110.3970 (accepted by Phys.\ Rev.\ D).

\bibitem{bib:sokoloff}
  A.~L.~Kagan and M.~D.~Sokoloff,
  {\it On Indirect $CP$ Violation and Implications for \Dz-\Dzb and \Bs-\Bsb mixing},
  Phys.\ Rev.\ D {\bf 80} (2009) 076008
  [arXiv:0907.3917].

\bibitem{bib:cdf_dacp}
  CDF Collaboration,
  T.~Aaltonen {\it et al.},
  {\it Measurement of $CP$-violating asymmetries in $D^0\to\pi^+\pi^-$ and $D^0\to K^+K^-$ decays at CDF},
  arXiv:1111.5023.

\bibitem{bib:confDACP}
  LHCb Collaboration,
  {\it A search for time-integrated $CP$ violation in $D^0 \rightarrow h^{+}h^{-}$ decays and a measurement of the $D^0$ production asymmetry},
  LHCb-CONF-2011-023.
\end{thebibliography}
\end{document}